\documentclass[preprint]{emulateapj}

\slugcomment{Submitted to the Astronomical Journal}
\shortauthors{Eskew et al.}
\shorttitle{}

\begin{document}
{\title{Converting From 3.6 and 4.5 $\mu$\lowercase{m} Fluxes to Stellar Mass}
  
\author{Michael Eskew$^{1,2}$, Dennis Zaritsky$^2$, and Sharon Meidt$^3$}
\affil{$^1$ Department of Astronomy, University of Texas, Austin, TX, 78712, USA}
\affil{$^2$ Steward Observatory, University of Arizona, 933 North Cherry Avenue, Tucson, AZ 85721, USA}
\affil{$^3$ Max-Planck-Institut f\"ur Astronomie, K\"onigstuhl 17, D-69117 Heidelberg, Germany}

\email{meskew@astro.as.utexas.edu, dzaritsky@as.arizona.edu, meidt@mpia.de}

\begin{abstract}    
We use high spatial resolution maps of stellar mass and infrared flux of the Large Magellanic Cloud (LMC) to calibrate a conversion between 3.6 and 4.5 $\mu$m fluxes and stellar mass, $ M_* =  10^{5.65} F_{3.6}^{2.85}F_{4.5}^{-1.85}{\left({D\over 0.05}\right)}^2 M_\odot$,
where fluxes are in Jy and D is the luminosity distance to the source in Mpc, and to provide an approximate empirical estimate of the fractional internal uncertainty in $M_*$ of  $0.3\sqrt{N/10^6}$, where $N$ is 
the number of stars in the region. We find evidence that young stars and hot dust contaminate
the measurements, but attempts to remove this contamination using data that is far
superior than what is generally available for unresolved galaxies resulted in marginal gains in accuracy. 
The scatter among mass estimates for regions in the LMC is comparable to that found by previous investigators when modeling composite populations, and so we conclude that our simple conversion is as precise as possible for the data and models currently available. Our results allow for a
reasonably bottom-heavy initial mass function, such as Salpeter or heavier, and moderately disfavor lighter versions such as a diet-Salpeter or Chabrier initial mass function.
\end{abstract}


\section{Introduction}
\label{sec:intro}

The measurement of the mass of the stellar population in a galaxy, $M_*$, is a direct, although integrated, measure of the star formation history of the system. We are increasingly realizing that $M_*$ is intricately connected to other galaxy properties, such as current star formation rate and morphology \citep{kauffmann}. Unfortunately, the measurement of $M_*$ is indirect and subject to significant systematic uncertainties. 

There are effectively two principal approaches to measure $M_*$. First, one can measure the dynamical mass of a galaxy, via kinematics \citep{cappellari} or lensing \citep{auger},
and then somehow model and subtract the contribution of dark matter to that measured mass. This approach is predicated on the successful subtraction of a large unseen mass, which is dominant in galaxies overall \citep{zaritsky94} and even substantial within the optical radius \citep{cappellari, auger}, and can easily lead to uncertainties of the order of the measurement itself.  Second, one can rely on stellar population models \citep[for example those of][]{bc} to connect $M_*$ to an observable, such as the luminosity in a selected passband, a color, or the 
spectral energy distribution as obtained either from spectroscopy or multi-band photometric observations. In this case, one is also subject to modeling uncertainties, although here those arise
from uncertainties in the stellar initial mass function (IMF), the stellar evolution models, particularly in the stages where stars are at their most luminous \citep[for some examples drawn from an extensive literature on the topic see][]{langer,maraston,conroy,mcquinn, dalcanton11}, and the star formation history. 
Careful use of scaling relations \citep{bell}, which are themselves indirect measurements of mass, 
provide a bridge across the two methods, greatly enlarge the sample size and help mitigate these problems, but result in prescriptions that may be accurate on average but highly uncertain in specific cases.

Having these two independent approaches to measure $M_*$ is advantageous in that the results can be compared to uncover any systematic differences. For example \cite{mclaughlin} compare the results obtained for stellar clusters and conclude that there are no egregious differences. The magnitude of the differences are now constrained to be relatively small (few tens of percent) and therefore modeling subtleties are critical and difficult
to control for.
We address this issue by presenting and applying a third method, which is more closely related to the photometric method outlined above but avoids several perilous assumptions. Our aim is to calibrate an easily observed quantity that is available for large numbers of galaxies and relatively impervious to the effects of extinction, ongoing star formation, and details of rare, but luminous phases of stellar evolution, and to  
determine the uncertainties introduced by
those phenomenon. We choose here the more modest goal of identifying and calibrating
such an estimator of $M_*$, due to our interest in applying such an estimator, rather than the more demanding goal of 
fully understanding the physical origin of the scatter and any possible systematic biases we uncover.

To help us measure $M_*$, 
we select the 3.6 and 4.5 $\mu$m fluxes, $F_{3.6}$ and $F_{4.5}$, which are becoming increasingly available for large
samples of galaxies \citep{sheth} with the advent of the {\sl Spitzer} \citep{werner} and {\sl WISE} \citep{wright} telescopes, and which are nearly minimally sensitive to young stellar populations and dust absorption and emission. We will test combinations of $F_{3.6}$ and $F_{4.5}$ as proxies for $M_*$ in one galaxy, the Large Magellanic Cloud (LMC), 
whose extinction and star formation history have been mapped in spatial detail \citep{zaritsky99,hz09} and that has been extensively observed with the {\sl Spitzer} telescope \citep{meixner}. 

The basic concept we exploit  is that 
we can use the 
star formation histories (SFHs) recovered from synthesizing the stellar optical color-magnitude diagrams  (CMDs) that are available for nearby galaxies \citep[e.g.][]
{hz04,hz09,dalcanton} to calculate the stellar mass on a region-by-region basis and use those measurements to calibrate $F_{3.6}$ and $F_{4.5}$ as tools with which to measure $M_*$. We then use the 
scatter in that correspondence to uncover any additional parameters that can be used to refine
the measurement and to determine the underlying uncertainty in this measurement. 
Although this approach shares some of the difficulties faced by  the stellar population synthesis approach (dependence on initial mass function and on stellar evolution models), it
eliminates the poorest constrained aspect of those models, the star formation history, and minimizes the effect of poorly understood rare phases of stellar evolution, such as the thermally pulsing AGB phase experienced by intermediate mass stars and the red, core He burning phase experienced by stars of masses $> 3.5 M_\odot$ \citep{melbourne}. The latter advantage is realized
because the CMD modeling depends on the {\it number} of such stars rather than their {\it luminosities}. 
These stars are relatively rare, and hence have little effect on the CMD modeling, but extremely 
luminous, and hence have a large effect on global colors and luminosities. 
Recently available SFH maps, as described in \S2.1, provide the necessary data to calculate resolved stellar mass maps, which are then compared to local measures of $F_{3.6}$ (\S2.2), to calibrate $F_{3.6}$ and $F_{4.5}$ as stellar mass tracers (\S3). We summarize our findings in \S4.

\section{The Input Data}
 
\subsection{Stellar Masses}

The spatially-resolved SFH of the Large Magellanic Cloud \citep[][hereafter HZ]{hz09}, provides the information necessary to construct a resolved stellar mass map of the galaxy for an adopted IMF. Because the regions are rectangular and the available algorithms in IRAF\footnote{IRAF is distributed by the National Optical Astronomy Observatories,
    which are operated by the Association of Universities for Research
    in Astronomy, Inc., under cooperative agreement with the National
    Science Foundation.}  to measure a luminosity work with circular apertures, we simply use the inscribed circle within each of the HZ regions to define the zones in which we will also measure the infrared luminosities, and correct for the differences in areas. The stellar mass we calculate is the integral over mass for the adopted IMF, Salpeter \citep{salpeter}, normalized by the star formation rate at a particular time,  integrated over time for the lifetime of the LMC. We will discuss later (\S3) a correction for the mass returned by evolved stars to the interstellar medium.

One technical point in this procedure regards the duration of the earliest (oldest) bin. Our analysis method \citep{hz01} finds the number of stars in the color magnitude diagram that match a population drawn from a particular isochrone, and then calculates a star formation rate by dividing that number by the length of time represented by that isochrone. That duration is easier to define when a particular isochrone is bracketed by other isochrones, but is not well defined for the oldest isochrone. As such, the star formation rate at earliest times will depend on the assumption of when the LMC started forming stars in this first age bin (e.g. did the LMC start forming stars 11 Gyr or 13 Gyr ago?). While this issue leads to some uncertainty in the derived star formation rate at these
times \citep[for example, see][]{mz} it does not pose a problem for the current analysis because we integrate over the length of the bin (we are interested here in the total
number of stars rather than in the rate of star formation). 

The uncertainties in the recovered star formation rates from CMD synthesis 
are difficult to ascertain \citep{hz01,dolphin,weisz}. Much of that uncertainty is related to how populations are partitioned among
adjacent bins in age or metallicity. These errors are unlikely to affect our results significantly because they arise from the fact that such populations are difficult to disentangle using broad-band photometry, and hence are unlikely to have highly differential properties in 
our photometry. Uncertainties over long times appear to be well controlled, as exemplified by tests done by \cite{hz01} using globular clusters, the global patterns in the star formation history seen in the LMC, and the correspondence of certain features in the star formation histories of the LMC and SMC \citep{hz04,hz09}. Such errors, even if they are significant, are likely to add randomly 
to the relationship between $M_*$ and $F_{3.6}$ because there is little
connection between stellar density and star formation history in the Clouds, outside of possibly in the LMC bar region \citep[see][]{hz09}. 

The key systematic uncertainty in our approach comes from the selection of the IMF. The behavior of the IMF at the high mass end ($M >$ few $M_\odot$) impacts the derivation of the SFH, as discussed for example by \cite{hz01}. In determining total stellar mass, the low mass form of the IMF is a potentially larger source of uncertainty. Here we assume a Salpeter IMF and note that scaling corrections corresponding to other choices of IMF can be calculated using 
population synthesis codes, such as P\'EGASE \citep{pegase}.

\subsection{IR Fluxes}

We use the published, calibrated mosaiced images of the LMC produced by the SAGE 
survey \citep{meixner} using the IRAC instrument  \citep{fazio} on the {\sl Spitzer} telescope to measure the fluxes, at both 3.6 and 4.5 $\mu$m, in 
circular apertures that are matched to the regions for which HZ present star formation histories.
The mosaics have pixels that are 2\arcsec on a side, or alternatively $9.4016 \times 10^{-11}$ steradians. The units of the surface brightness measurements are MJy/sr. 
We use the IRAF task PHOT, with a background value set to zero to calculate the flux in the region.  The majority of the regions are quite luminous, so the uncertainties are dominated by Poisson statistics and much smaller than those associated with 
the calculation of the stellar mass. After integrating over an aperture and converting units, 
we then present fluxes in Jy.

\begin{figure}
\plotone{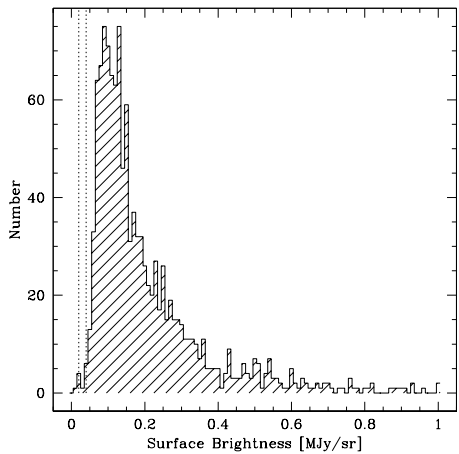}
\caption{The distribution of $F_{3.6}$ measurements. The sharp drop in distribution near a value of 0.04, shown by the right, dotted, vertical line, indicates our
upper limit on a possible uniform sky value (see text for discussion). Our preferred sky value of 
0.02, shown by the left vertical line, falls close to values obtained for the darkest regions available.}
\label{fig:aphist}
\end{figure}

As mentioned previously, we have artificially set the background to zero. Now we attempt to 
estimate the background directly from our results.  Our challenge is that the LMC extends beyond the region covered by the mosaic, so there are no ``empty" regions within the image. First, we examine the distribution of measured mean surface brightnesses within the various apertures (Figure \ref{fig:aphist}). We do this rather than examine the image directly, searching for dark patches, because we are not attempting to estimate the blank field background but rather the mean contribution from the blank field plus galactic stars plus any galactic diffuse emission. As such, selecting the darkest regions could result in a background estimate that is biased low.
From Figure \ref{fig:aphist}, we conclude that a robust upper limit to the mean background is likely to be 0.04 MJy/sr, at which point the number of apertures drops precipitously, signaling the lower limit of LMC-related emission. A slight peak in values at 0.02 MJy/sr suggests an alternate choice for the correct background level. 

\begin{figure}
\plotone{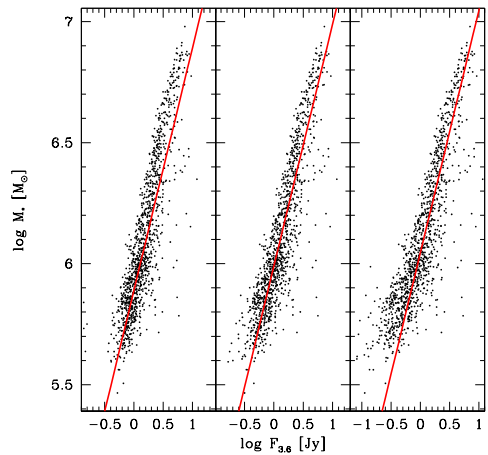}
\caption{An alternative justification for our choice of background level. The left panel shows the relationship between $F_{3.6}$ and $M_*$ when no background correction is applied. The middle panel shows the result with our preferred background value of 0.02 MJy/sr. The right panel shows the result when adopting the upper limit of 0.04 MJy/sr. When no correction is applied, the lower part of the distribution bends slightly downward below the best fit line. When our upper limit correction is applied, we see a tail of points that suggests an unphysical, asymmetric increase of the scatter for regions with lower total stellar mass.  Our preferred value minimizes the veritcal scatter about the best fit line (the horizontal scatter due to measurement errors is negligible given the large count values in these regions).}
\label{fig:skycomp}
\end{figure}

In Figure \ref{fig:skycomp}, where we plot the relationship between $F_{3.6}$ and $M_*$, we show the results of a different approach to estimating the background. 
We calculate the vertical scatter about the best fit linear relationship for different adopted background values and identify the background level that minimizes the scatter. The horizontal scatter introduced by measurement errors is negligible given the large fluxes over these regions. We find that the best fit value is 0.02 MJy/sr. 
In the three panels, we show the relationship for adopted background values of 0.00, 0.02, and 0.04 MJy/sr. If no background correction is applied, then there is a slight deviation of the lower end of the distribution away from the 1:1 line. The lack of a strong turnover in the relationship demonstrates that the background term is not highly significant. When we adopt our upper limit of 0.04 MJy/sr, we oversubtract the background contribution in many fields leading to the leftward tail of regions with  significant stellar mass but little IR emission. For our preferred value, 0.02, neither of these problems occurs. A few regions asymmetrically scatter to the right in the Figure and there is a systematic deviation from the
relationship at large flux values, but we will discuss the origin of those issues below. We do the same
analysis for $F_{4.5}$ and find a best-fit background of 0.017 MJy/sr.  These
backgrounds are subtracted from all fluxes discussed subsequently.
 
\section{Results and Discussion}
 
We present the results of our analysis of the fluxes and stellar masses for each LMC region in
Figure \ref{fig:colcorr}. The upper left panel in the Figure shows the raw result, using our adopted background value, for the correlation between $F_{3.6}$ and $M_*$. The solid line represents the mean linear relationship (slope $=$ 1 in our log-log plots). The dispersion in values about this line is 0.121 dex, when we
exclude points that deviate from the line by $>$ 0.3 dex (shown by the dotted red lines), and
the distribution of residuals is reasonably well approximated by a Gaussian. This degree
of scatter corresponds to mass estimates that have 1$\sigma$ uncertainties of $\sim$ 30\%. 
 
\begin{figure}
\plotone{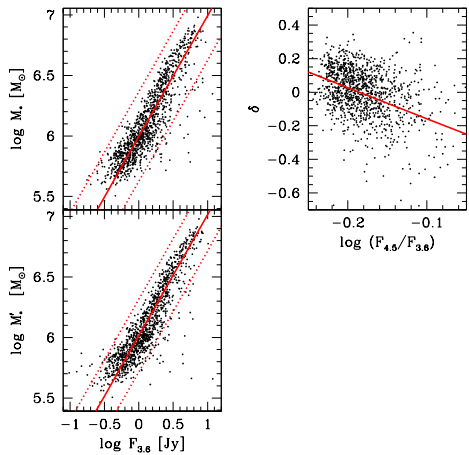}
\caption{The dependence of the relationship between $F_{3.6}$ and $M_*$ on IR color.  The upper left panel reproduces the relation shown in the middle panel of Figure \ref{fig:skycomp}. The upper right panel shows the residuals from the best fit line in the upper left panel, $\delta$, as a function of IR color (slope $-1.85$) The lower left  panel shows the relationship between $F_{3.6}$ and $M_*$ once we correct for the relationship in the residuals with color.  The dotted lines are $\pm 0.3$ from the best fit and represent the region used to evaluate the dispersions quoted in the text.}
\label{fig:colcorr}
\end{figure}

This scatter can, in principle, be reduced if it arises from physical sources, and if we can identify 
and account for those sources.
One such potential source is the variation in star formation history from region to region across the LMC. As such, one might suspect that using an independent measure of SFH variations, or mean stellar age, to refine the relationship between $F_{3.6}$ and $M_*$ would lead to lower scatter. We investigate utilizing the one simple directly observable measurement we have of such population variations, color, in Figure \ref{fig:colcorr}. 
In the upper right panel, we show the residuals in the original relationship, $\delta$, vs. color, $\log (F_{4.5}/F_{3.6}$). A linear relationship with slope $-1.85$ is shown in red and suggests that there is indeed a weak
connection between the residuals and this measure of the stellar populations. When we account for this correlation between IR color and residual, there is modest improvement (lower left panel Figure \ref{fig:colcorr}) and a dispersion about the best fit line of 0.117, similar to before the correction, but the systematic deviation of the upper end of the distribution relative to the lower
end is lessened. 
 
\begin{figure}
\plotone{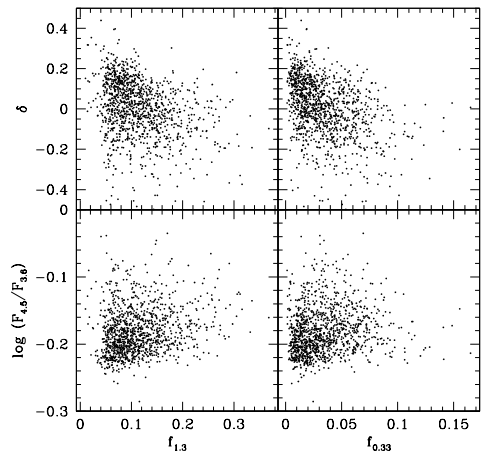}
\caption{Correlations between the residuals from the mean relation between $F_{3.6}$ and $M_*$,
 $\delta$, and the fraction of stars younger than 1.3 Gyr (upper left) or 0.33 Gyr (upper right), and between IR color and these same fractions (lower panels).  All panels contain statistically significant correlations, but the strongest is between $\delta$ and  $f_{0.33}$ (see text for correlation values).}
\label{fig:agecomp}
\end{figure}

Given the data we have for the LMC, which we will not have in general, we can probe somewhat further into the role of stellar populations in affecting $F_{3.6}$ as a mass tracer. 
Although we will not be able to apply this knowledge in general, the exercise might help highlight
the underlying cause for the scatter and provide guidance for other approaches aimed at dealing with
variations in $M/L_{3.6}$ \citep{meidt}. In Figure \ref{fig:agecomp} we show the correlations between the residual, $\delta$, and the fraction of the
stellar population, $f$, that is younger than 1.3 Gyr (upper left panel) and 0.33 Gyr (upper right panel). 
In the lower panels of the Figure, we show the relationship between those same fractions and
the IR color, which we
used previously to correct the relationship. The correlations
between $\delta$ and either $f_{1.3}$ or $f_{0.33}$ are both strong, with the stronger being with the youngest stellar fraction (Spearman rank correlation coefficients of $-0.328$ and $-$0.493 respectively). Both of these correlation coefficients correspond to 
highly significant correlations, with probabilities of occurring randomly in a sample of this size $< 6 \times 10^{-31}$, and they are both stronger than the correlations seen with color (correlation coefficients of 0.177 and 0.118, respectively). Interestingly, the residual from the mean trend of 
$F_{3.6}$ vs. $M_*$ is a better diagnostic of the fraction of very young stars than is the IR color. The finding that $f_{0.33}$ is more strongly correlated
 with $\delta$ than color is with $\delta$ ($-0.493$ vs. $-$0.348) demonstrates that IR color
 alone cannot act to correct the relationship entirely for stellar population variations. 
 Importantly, even relatively modest
 fractions of young stars ($\sim $5\%) are sufficient to significantly affect $M_*/F_{3.6}$. 
 
Although we have found that the residuals in the $F_{3.6}-M_*$ relationship correlate with
$f_{0.33}$ and that IR color cannot fully correct for this trend, we find that  correcting for
the trend does not significantly lower the scatter in the $F_{3.6}-M_*$ relationship. 
We correct for the correlation between $\delta$ and $f_{0.33}$ by fitting a linear relationship
between $\delta$ and log$(f_{0.33}$) to produce the results shown in Figure \ref{fig:agecorr}.
The corrected data have a scatter of 0.115 (30\% in mass). Some improvement is
visible, particularly in the $\delta$ distributions (lower panels), but the decrease in the scatter is 
marginal. We conclude that using even superior data than a single IR color to remove 
stellar population variations (at the level of a linear fit to the residuals) is insufficient to 
substantially lower the scatter. Furthermore, this level of scatter is comparable to that inherent to more involved and data-intensive treatments, such as the two-color method at H-band developed by \cite{zibetti}.

\begin{figure}
\plotone{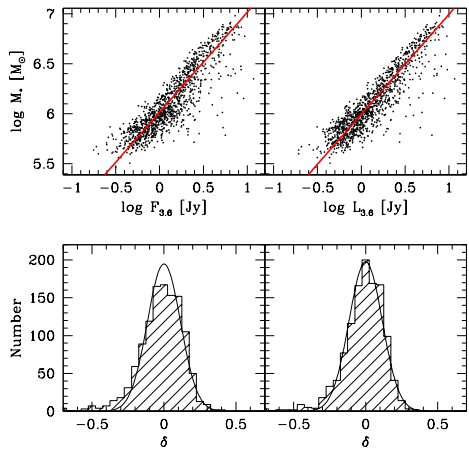}
\caption{The dependence of the relationship between $F_{3.6}$ and $M_*$ on the fraction of stars younger than 0.33 Gyr, $f_{0.33}$. In the left panels we plot the uncorrected relationship between $F_{3.6}$ and $M_*$ and the distribution of residuals, $\delta$ about that relationship. In the right panels we plot the same, except here we have applied a best fit correction between $\delta$ and $f_{0.33}$. Although improvement in the relation is evident, 
it is relatively modest even though we have a direct measurement of the fraction of young stars.}
\label{fig:agecorr}
\end{figure}

\begin{figure}
\plotone{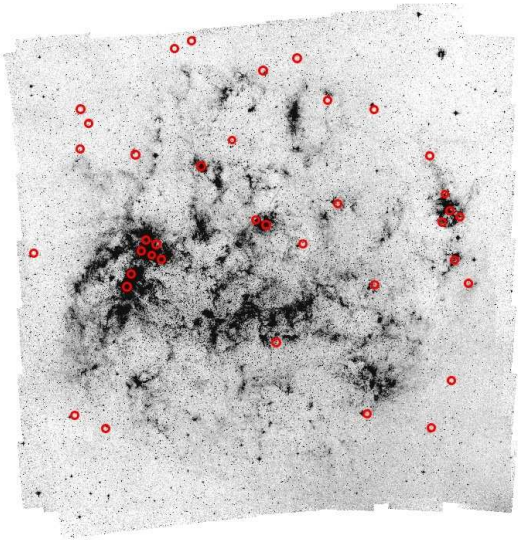}
\caption{Spatial distribution of regions with large residual from the $F_{3.6}$ vs. $M_*$ relationship, superposed on the 8 $\mu$m mosaic image of the LMC from the SAGE survey \citep{meixner}. Specifically, we selected regions with residuals $< -0.3$ from the age-corrected
relationship shown in Figure \ref{fig:colcorr}. Although many of the outliers correspond to regions
of high 8 $\mu$m flux, about a third are in regions that appear to be unremarkable.}
\label{fig:8mu}
\end{figure}

In addition to the issue of scatter about the mean relationship, 
there is a population of regions that consistently lie
significantly rightward of the mean relation between $M_*$ and $F_{3.6}$ (Figures \ref{fig:colcorr}
and \ref{fig:agecorr}). 
To understand the origin of this discrepancy, we plot the location of regions
within the LMC with
large residuals  ($\delta < -0.3$) on the 8 $\mu$m image of the LMC (Figure \ref{fig:8mu}), 
also from the SAGE survey \citep{meixner}. We find that many of these outliers correlate with regions of hot dust, as traced by the 8 $\mu$m image. However, this correspondence is far from 
ideal and some outliers lie in unremarkable regions, highlighting the difficulty in using any color, even one that identifies regions with warm dust, to correct $F_{3.6}$ for contamination. The contamination probably comes from various components, ranging from wide-scale hot dust
that is easily visible in the image to compact sources such as
 extreme AGB stars, which can have $J - 3.6$ colors in excess of 5 \citep{blum}. While the latter
 are evident when one has photometry of individual stars, as can be obtained in the LMC \citep{blum}, they will be difficult to pick up in integrated populations, in data with limited color baselines, and in data with limited S/N over small spatial scales.
 Even with sophisticated treatments, such as that described by \cite{meidt}, one might at best
 be able to exclude {\sl all} of these outliers from the analysis, at which point one would return to our
 estimate of the scatter of $\sim$ 0.12 (or 30\% in mass for the regions considered here). 

One aspect that we have neglected so far is the return of mass to the interstellar medium by evolved stars. What we have calculated is the sum of the mass of all stars formed, which is different than the sum of the mass of all current stars. To estimate the systematic difference between the two, we use P\'EGASE models \citep{pegase} with the global SFH of the LMC as we did in \cite{mz}. We find that 94\% of all the mass formed into stars by the current time remains in stars and remnants. We will therefore correct our calculated stellar masses by 0.94. In principle, this correction should vary with the SFH of each region, but the differences are well below the 30\% region-to-region scatter, and so we neglect them.
  
The color-corrected relationship shown in Figure \ref{fig:colcorr} suggests that the remaining intrinsic uncertainty, excluding the asymmetric tail, is about 0.12 dex, or about 30\% for regions containing somewhere in the vicinity of 10$^{5.5}$ to 10$^7 M_\odot$. If this scatter is random, then estimating the masses for larger regions should result in lower fractional uncertainty. We can estimate how well
these fluctuations average out by combining all of our data into an estimate of $M_*$ for the entire
LMC and comparing to the extrapolation of our 1:1 calibrated relationship. To do this, we exclude
regions that have $\delta < -0.3$. We find that result is exceedingly close to the extrapolated relation (using only $F_{3.6}$ we infer $M_* = 1.72\times 10^9 M_\odot$ and we
calculate $M_* = 1.81 \times 10^{9} M_\odot$ using the stellar masses for 1131 out of 1180 regions and the correction for mass returned to the ISM). If we do not remove any outliers, we obtain $1.90\times10^9$ and $1.87\times10^9 M_\odot$, respectively, and similar results ($1.81 \times 10^9$ and $1.93\times 10^9 M_\odot$, respectively)
 if we use the relationship with both $F_{3.6}$ and $F_{4.5}$. Either way, 
this analysis implies that the internal uncertainty drops to a few percent when considering systems with $M_* \sim 2\times 10^9 M_\odot$. This gain of a factor of at least 10 in precision corresponds well to the naive $\sqrt{N}$ gain expected in going from regions with $\sim 10^6$ to $\sim 10^{9} M_\odot$. We conclude
that scaling the uncertainty by $\sqrt{N}$ provides a reasonable estimate of the expected gain (or loss) in precision and suggest that on galaxy scales we are limited by systematic rather than  statistical uncertainties.

One final check of our results is possible by comparing the dynamical mass, $M_D$, of the LMC out
to the radius sampled here and the stellar mass calculated above. For a rotation speed of 
 87 km s$^{-1}$ \citep{olsen} out to a radius of 3 kpc, corresponding to the area
covered by our survey, we estimate that the enclosed mass 
is $5.0\times10^9$ M$_\odot$. Given the unknown contributions of gas and dark matter, plus the $o(1)$ geometric corrections necessary in going from our assumed spherical geometry in using $M = rv^2$ to a more realistic disk potential, the two estimates are not in significant conflict, and certainly satisfy the basic constraint that $M_* < M_D$. 

Our preferred calibrations at the distance of the LMC are therefore 
$$ M_* = 10^{5.97} F_{3.6}$$
when $F_{4.5}$ is unavailable and 
$$ M_* =  10^{5.65} F_{3.6}^{2.85}F_{4.5}^{-1.85}$$
otherwise, where $F_{3.6}$ and $F_{4.5}$ are fluxes in Jy and $M_*$ is given in solar masses. The expression for a source at arbitrary distance, $D$, in Mpc is
$$ M_* =  10^{5.65} F_{3.6}^{2.85}F_{4.5}^{-1.85}{\left({D\over 0.05}\right)}^2$$
if we adopt a distance of 50 kpc for the LMC. 

Previous work combining stellar mass estimates from integrated spectroscopy and photometry \citep{zhu} provided an alternative calibration of the relationship between $F_{3.6}$ and $M_*$. They present two transformations, one that depends solely on $L_{3.6}$ and another that also utilizes $g-r$. We can only apply the former because we do not have a measure of the $g-r$ color of the LMC. To apply their relationship, we need to convert our measured $F_{3.6}$, which is 2012 Jy  for the the sum of all of the regions we studied, to $L_{3.6}$ in solar units. We use the zero point IRAC channel 1 calibration provided by \cite{reach} (0 Vega magnitude corresponds to 280.9 Jy) and a 3.6 $\mu$m magnitude of the Sun of 3.24 \citep{oh} to calculate that $L_{3.6}$ for the LMC is $3.7 \times 10^{9} L_\odot$. Applying their Eq. 2 results in an estimate of $M_*$ of $4.0 \times 10^{10} M_\odot$, which is a factor of eight larger than the dynamical mass. One potential source of the discrepancy is the choice of the Solar magnitude (because they do not quote their adopted value), although that seems unlikely to create an eight-fold difference. Another potential source is our measured $L_{3.6}$, which could either be affected by errors or by contamination of luminous sources. We checked against errors by measuring the luminosity from a single aperture across the image (rather than summing the results from all of our subregions) and by comparing the luminosity to the 3.6 $\mu$m Tully-Fisher relation provided by \cite{freedman}. Both comparisons confirm that there is no significant problem with our measured luminosity. The problem may lie with extrapolating the fitted \cite{zhu} relation to the LMC, because adopting reasonable $g-r$ colors for the LMC does decrease the discrepancy somewhat (although it does not bring down the estimated $M_*$ below $M_D$) or may reflect a metallicity dependence, which they acknowledge to have not addressed. 

Finally, we close with a discussion of the gap between $M_D$ and $M_*$, which is substantial ($\sim 3 \times 10^9 M_\odot$ or a factor of $\sim$ 2.5). This ``shortfall" could be ascribed to the ubiquitous dark matter, but it also leaves room for significant amounts of baryonic matter.
We focus now on the impact of the choice of the IMF and implications for the study of unresolved stellar populations.  There continues to be much debate over the appropriate functional form of the IMF and whether it is variable either as a function of  galaxy type \citep{vandokkum} or across redshift \citep{dave}. There are two principal difficulties in applications of the stellar population models in constraining the IMF in systems with unresolved stellar populations. First, the population of stars that dominate the total luminosity typically includes evolved stars, which are notoriously difficult to model. The measured luminosity, relative to the model one, sets the normalization of the IMF. Second, the bulk of the inferred mass comes from low mass stars, which are not observed, leading to the strong sensitivity of the implied mass on the form of the IMF. \cite{vandokkum}, using a spectral signature of low mass stars, recently argued (for giant elliptical galaxies) that the slope of the IMF might be significantly steeper than Salpeter, leading to a vast reservoir of unappreciated stellar mass. Our study by-passes the first problem by counting stars rather than integrating luminosity, but is susceptible to the second one. For reference in comparing the effects of IMF choice, an IMF favored by \cite{bell} called the diet-Salpeter, because it reduces the total integrated stellar mass by limiting the number of low mass stars relative to Salpeter, results in 0.1 dex less implied mass, while the use of the \cite{chabrier} mass function results in 0.25 dex less implied mass. 

Our calculations above result in an inferred $M_*/L_{3.6} = 0.5$ for the LMC $(1.9 \times 10^9 M_\odot/ 3.7 \times 10^9 L_\odot)$. This is already a fairly low value of $M_*/L_{3.6}$, which would only decrease further if either the diet-Salpeter or Chabrier IMFs were adopted. Given that the dynamical mass is already a factor of $\sim$ 2.5 higher than the stellar mass, we contend that lowering the stellar mass even further is disfavored because we do not expect the dark matter fraction to be significantly larger than 60\% in the inner 3 kpc of the LMC \citep{alves}.  For example, adopting a Chabrier IMF (applying simple the 0.25 dex correction rather than re-calcuating the SFH) results in the dark matter fraction increasing to $\sim $ 80\%. Such a high dark matter fraction is at the limit of the range found necessary to produce the baryonic Tully-Fisher relation \citep{mcgaugh}, and it is found in galaxies of lower masses and surface brightnesses than the LMC. Going in the other direction, it is interesting to consider the bottom-heavy IMF suggested for giant elliptical galaxies by \cite{vandokkum}. To fully consider the effects of an IMF that is different than Salpeter at all stellar masses, rather than for stars with low mass that are not constrained by \cite{hz09}, requires recalculating the SFH with this new IMF. However, if we simply calculate the difference in stellar mass between Salpeter ($x = - 2.35$) and the IMF suggested by \cite{vandokkum} ($x = -3$) when we normalize the two to have the same number of stars with $M > 1 M_\odot$), we find that the bottom-heavy IMF has a factor of 2.5 more mass (for stars with $0.1 < M < 100 \ M_\odot$). Given that our dynamical mass estimate is 
$5 \times 10^{10}M_\odot$, we find that such a steep IMF is consistent with our measurements, but only if there is a negligible dark matter fraction over the inner 3 kpc of the LMC. 

In contrast, various other studies of late type galaxies contend that bottom heavy IMFs are inconsistent with dynamical or lensing mass constraints \citep[see, for example,][]{brewer}.
Given the excellent data and analysis in the \cite{brewer} study, the only way we can envision reconciling the results is
if the stellar population models are significantly underpredicting the luminosities used to 
match the observed luminosities. 
If the stellar models do underpredict ithe luminosity, that would in turn lead investigators to normalize the IMF's too high, resulting in a mass that violated dynamical constraints, unless the IMF
is constructed to significantly turn over at low masses. Intriguingly, some modelers are reaching analogous conclusions for independent reasons \citep{leitherer}. Alternatively, a systematic error lurks in our calculation of the stellar mass corresponding to the published SFH. Resolving this issue is manifestly paramount to many areas of astrophysics.

\section{Summary}

We have used spatially resolved maps of stellar mass and IR flux in the Large Magellanic Cloud to 1) calibrate a conversion between 3.6 and 4.5 $\mu$m fluxes and stellar mass, 2) examine potential approaches for using either 4.5 $\mu$m or 8 $\mu$m data to refine the estimates of stellar mass, and 3) provide empirical estimates of the uncertainty in such measurements. 

We find that:

\medskip
\noindent
One can use measurements of the fluxes at 3.6 and 4.5 $\mu$m, $F_{3.6}$ and $F_{4.5}$, to estimate the corresponding
stellar mass using
$$ M_* =  10^{5.65} F_{3.6}^{2.85}F_{4.5}^{-1.85}{\left({D\over 0.05}\right)}^2$$
where $M_*$ is in solar masses, $F$'s are in Jy, and D is the distance to the source in Mpc.

\medskip
\noindent
Although we were able to clearly identify deviations from the mean relationship between
flux and stellar mass that correlate with stellar population variations, in particular with populations $<$
300 Myr old, and with hot dust, as traced by 8 $\mu$m emission, correcting for those sources
of scatter, with data far superior to what will generally be available for most galaxies, results
in marginal reductions in the scatter. Furthermore, we find that small fractions of such young populations ($<$ 5\%) are sufficient to significantly affect $F_{3.6}$. We conclude that it is difficult to reduce the scatter below what
we find using broad, global measurements.

\medskip
\noindent
The scatter, for our regions, which typically contain between $10^{5.5}$ and $10^{6.5} M_\odot$ (and hence we presume comparable numbers of stars), is approximately 30\% in mass. This scatter decreases as $\sqrt{N}$ as regions with more stars are analyzed, resulting in an estimate of $M_*$ that has an internal precision of $<$ a few percent for the entire LMC. 

Although the LMC contains a range of environments, covering strongly star-forming regions such as 30 Dor to  quiescent regions, the relationship we provide may break down for even more strongly star-forming systems. We also have not explored the dependence of this calibration on metallicity. Within the LMC there is little variation in metallicity \citep{pagel}, making it difficult to explore this issue
within the context of this study. However, one could extend this work to the SMC and to more distant galaxies for which resolved CMDs exist to establish the magnitude of the metallicity dependence. Finally, we have not
directly addressed the uncertainty resulting from the adopted IMF. However, because this is a systematic uncertainty it will be more prominent in some uses of the relation, such as when total stellar masses are needed, than in other uses, such as when one is studying the distribution of stellar mass within one system. We have argued that our results can accommodate the conjecture of a relatively bottom-heavy IMF for the LMC, such as Salpeter, and are in moderate conflict with  bottom-light IMFs such as the 
diet-Salpeter and Chabrier, which are usually advocated for late-type galaxies.
Bearing these caveats in mind, we expect our calibrated conversion to be particularly useful in analyzing the large amount of extragalactic {\sl Spitzer} and {\sl WISE} data already in hand. 

\begin{acknowledgments}

We acknowledge financial support from NASA LTSA  award 
NNG05GE82G and NSF grants AST-0307482 and AST-0907771.

\end{acknowledgments}


\begin{thebibliography}{}

\bibitem[Alves \& Nelson(2000)]{alves}
Alves, D.R., \& Nelson, C.A. 2000, \apj, 542, 789

\bibitem[Auger et al.(2006)]{auger}
Auger, M.W., Treu, T., Bolton, A.S., Gavazzi, R., Koopmans, L.V.E., Marshall, P.J., Bundy, K., \& Moustakas, L.A., 2009, \apj, 706, 1099

\bibitem[Bell et al.(2004)]{bell}
Bell, E. F. et al. 2004, 608, 752

\bibitem[Blum et al.(2006)]{blum}
Blum, R., et al. 2006, \aj, 143, 2034

\bibitem[Brewer et al.(2012)]{brewer}
Brewer, B.J. et al. 2012, \mnras, in press.

\bibitem[Bruzual \& Charlot(2003)]{bc}
Bruzual, A. G. , \& Charlot, S. 2003, \mnras, 344, 1000

\bibitem[Cappellari et al.(2006)]{cappellari}
Cappellari, et al. 2006, \mnras, 366, 1126

\bibitem[Chabrier(2003)]{chabrier}
Charbrier, G. 2003, PASP, 115, 763

\bibitem[Conroy et al.(2009)]{conroy}
Conroy, C., White, M., \& Gunn, J.E. 2010, \apj, 708, 58

\bibitem[Dalcanton et al.(2009)]{dalcanton}
Dalcanton, J.J., et al. 2009, \apjs, 183, 67

\bibitem[Dalcanton et al.(2011)]{dalcanton11}
Dalcanton, J.J., et al. 2011, \apjs, in press

\bibitem[Dav\'e(2008)]{dave}
Dav\'e, R. 2008, \mnras, 385, 147

\bibitem[Dolphin(2002)]{dolphin}
Dolphin, A.E. 2002, \mnras, 332, 91

\bibitem[Eskew \& Zaritsky(2011)]{mz}
Eskew, M., \& Zaritsky, D. 2011, \aj, 141, 69

\bibitem[Fazio et al.(2004)]{fazio}
Fazio, G.G. et al. 2004, \apjs, 154, 10

\bibitem[Fioc \& Rocca-Volmerange(1997)]{pegase}
Fioc, M., \& Rocca-Volmerange, B. 1997, A\&A, 326, 950

\bibitem[Freedman et al.(2011)]{freedman}
Freedman, W., et al. 2011, \aj, 142, 192

\bibitem[Gordon et al(2011)]{gordon}
Gordon, K., et al. 2011, \aj, submitted

\bibitem[Harris \& Zaritsky(2001)]{hz01}
Harris, J., \& Zaritsky, D. 2001, \apjs, 136, 25

\bibitem[Harris \& Zaritsky(2004)]{hz04}
Harris, J., \& Zaritsky, D. 2004, \aj, 127, 1531

\bibitem[Harris \& Zaritsky(2009)]{hz09}
Harris, J., \& Zaritsky, D. 2009, \aj, 138, 1243

\bibitem[Kauffmann et al.(2004)]{kauffmann}
Kauffmann, G., White, S.D.M., Heckman, T.M., M\'enard, B., Brinchmann, J., Charlot, S., Tremonti, C., \& Brinkmann, J. 2004, \mnras, 713, 731

\bibitem[Langer \& Maeder(1995)]{langer}
Langer, N., \& Maeder, A. 1995, A\&A, 295, 685

\bibitem[Leitherer \& Ekstr\"om(2011)]{leitherer}
Leitherer, C., \& Ekstr\"om, S. 2011, IAU Proceedings, 284, 2011

\bibitem[Maraston et al.(2006)]{maraston}
Maraston, C., Daddi, E., Renzini, A., Cimatti, A., Dickinson, M., Papovich, C., Pasquali, A., \& Pirzkal, N., 2006, \apj, 652, 85

\bibitem[McGaugh(2005)]{mcgaugh}
McGaugh, S.S. 2005, \apj, 632, 859

\bibitem[McLaughlin \& van der Marel(2005)]{mclaughlin}
McLaughlin, D.E., \& van der Marel, R.P. 2005, \apjs, 161, 304

\bibitem[McQuinn et al.(2011)]{mcquinn}
McQuinn, K.B.W., Skillman, E.D., Dalcanton, J.J., Dolphin, A.E., Holtzman, J., Weisz, D., \& Williams, B.F. 2011, \aj, 740, 48

\bibitem[Melbourne et al.(2012)]{melbourne}
Melbourne, J. et al. 2012, \apj, 748,47

\bibitem[Meidt et al.(2012)]{meidt}
Meidt, S., et al. 2012, \apj, 744, 17

\bibitem[Meixner et al.(2006)]{meixner}
Meixner, M., et al. 2006, \aj, 132, 2268

\bibitem[Oh et al.(2009)]{oh}
Oh, S.-H., de Blok, W. J. G., Walter, F., Brinks, E., and Kennicutt, R.C., Jr. 2008, AJ, 136, 2761

\bibitem[Olsen et al.(2011)]{olsen}
Olsen, K., Zaritsky, D., Blum. R.D., Boyer, M.L. and Gordon, K.D. 2011, \apj, 737, 29

\bibitem[Pagel et al.(1978)]{pagel}
Pagel, B.E.J., Edmunds, M.G., Fosbury, R.A.E., \& Webster, B.L. 1978, \mnras, 184, 569

\bibitem[Reach et al.(2005)]{reach}
Reach, W.T., et al. 2005, PASP, 117, 978

\bibitem[Salpeter(1955)]{salpeter}
Salpeter, E.E. 1955, \apj, 121, 161

\bibitem[Sheth et al.(2010)]{sheth}
Sheth, K. et al. 2010, PASP, 122, 1397

\bibitem[Tully \& Fisher(1977)]{tf}
Tully, R.B., \& Fisher, J.R. 1977, A\&A, 54, 661

\bibitem[van Dokkum \& Conroy(2010)]{vandokkum}
van Dokkum, P. G., \& Conroy, C. 2010, Nat, 468, 940 

\bibitem[Werner et al.(2004)]{werner}
Werner, M.W., et al. 2004, \apjs, 154, 1

\bibitem[Weisz et al.(2011)]{weisz}
Weisz, D.R., et al. 2011, \apj, 739, 5

\bibitem[Wright et al.(2010)]{wright}
Wright, E.L. et al., 2010, \aj, 140, 1868

\bibitem[Zaritsky(1999)]{zaritsky99}
Zaritsky, D., 1999, \aj, 118, 2824

\bibitem[Zaritsky \& White(1994)]{zaritsky94}
Zaritsky, D., \& White, S.D.M. 1994, ApJ, 435, 599

\bibitem[Zibetti et al.(2009)]{zibetti}
Zibetti, S., Charlot, S., \& Rix, H.-W. 2009, MNRAS, 400, 1181

\bibitem[Zhu et al.(2010)]{zhu}
Zhu, Y.-N., Wu., H., Li, H.-N., and Cao, C. 2010, RAA, 10, 329

\end{thebibliography}
\end{document}